# MMOGs as Social Experiments: the Case of Environmental Laws

Joost Broekens

[1] Telematica Instituut, Enschede, The Netherlands
Joost.broekens@telin.nl

**Abstract.** In this paper we argue that Massively Multiplayer Online Games (MMOGs), also known as Large Games are an interesting research tool for policy experimentation. One of the major problems with lawmaking is that testing the laws is a difficult enterprise. Here we show that the concept of an MMOG can be used to experiment with environmental laws on a large scale, provided that the MMOG is a real game, i.e., it is fun, addictive, presents challenges that last, etc.. We present a detailed game concept as an initial step.

## 1 Introduction

Massively Multiplayer Online Games (MMOGs; see [1]), also known as Large Games [4] are computer games in which large numbers of players share an online virtual world in which they act to achieve certain game-related goals. Key elements in an MMOG are that the environment is online, it has a large number of active players involved (typically over 100.000) and that it is a game, meaning the goal for a player to interact in the virtual world is entertainment and the virtual world has built-in game-rules (see also [2, 4]). For comparison, Second Life, a virtual world in which persons can create a form of virtual existence but lacks game elements in its design is not a game and therefore not a MMOG.

In this paper we argue that MMOGs can be good research tools for policymakers, especially if the game is designed for this purpose. This is in line with work by Castranova [4], Bradley and Froomkin [2] and Daalen et al [5]. Here we first argue why MMOGs have this characteristic. Second, we address energy and pollution laws as a potential application of MMOGs as a policy maker's research tool. Finally we present the (at present non-existing) MMOG *Energetics* and detail how the game could be designed. From a research perspective this game enables large-scale experimenting with new environmental law systems in order to see how players cope with these laws. As such, *Energetics* is not only a game, it can be used as a tool to help policy making by simulation.

## 2  Why are MMOGs serious even as a game?

The entertainment computing literature often makes a distinction between computer games that are meant to be serious (such as training simulations), i.e., serious gaming, and games that are meant to be fun, i.e. games. This does not exclude the possibility that a serious game is fun, but it somehow gives a priority. This is especially true in the field of games that promote awareness or are designed to educate. Such games are often not developed by the gaming industry but by researchers and policy makers. The result is that many of these games are actually not fun to play, and they are in essence a re-packaged educational message. This approach will not work when MMOGs are being used for social experiments. The MMOG has to be designed as a game, not as a "message". In other words, it has to be fun and it must have realistic MMOG challenges such as character advancement, socializing, exploring, competition, teamwork, etc. [10, 11].

An MMOG can be made serious while being a game in two main ways. First, the game goals can be made serious. Second the game mechanisms can be made serious. The first refers to what kind of things players can achieve in the game. For example, in a game aimed at promoting multiracial integration in the real world, the game designers could build quests (playable storylines) that have as goal to become friends with another in-game race by understanding their culture in the game world. The second refers to how the game works. For example, the market system in the science fiction MMOG *EVE Online* (www.eve-online.com; *CCP games*) is comparable to a real market with supplies, demands, production, inflation, etc. Further, buyers need to pick up a bought item at the spot where the seller has dropped it. This results in phenomena such as the emergence of vivid market locations in the game (e.g., the Solar system called *Jita*; see also [5] on the topic of MMOG market analysis).

Even when an MMOG is fun and has some seriousness to it, this of course does not necessarily mean it is a useful research tool. Issues arise such as "is the player base a good sample of the population" (demographics wise), and, "do players behave in a way that is comparable to their behavior in the real world". In this paper we do not address the first issue, as this is strongly dependent on the actual content of the game, the research question that one wants to address with the game, as well as general gaming trends (such as increased numbers of women that play computer games). This is a full topic on its own. The second issue essentially boils down to: how do choices of players in the game world reflect their choices in similar situations outside the game world? This question is not answerable in a generic way: in some games for example, the choices you have in the game world do simply not exist in the real world. We present four indications to back up our claim that MMOG's are serious even when essentially being a game and that the players' choices reflect real-world choices considerably.

First, consider again the market system in *EVE Online*. As mentioned earlier, because items have to be "shipped" (i.e., they do not magically move to the buyers inventory), actual market hotspots arise that are in locations that are easily reachable from through the galaxy (the *EVE Online* world is a large galaxy with 6000+ solar systems). Also, there is a very active trade profession. People move items from one spot to another because the can make a good "virtual buck". This is also due to the complex but realistic item production mechanisms. Some people make items cheap, because they invested a lot of effort in training their production skills. Due to the game mechanics that have been chosen for the EVE market and item production system, we can see that the *EVE Online* economy is healthy and is compatible with how markets work in the real world [9]. This means that game mechanics can influence gamers' behavior such that it becomes plausible from a real world perspective.

A second indication is the fact that in many MMOGs investment in your character is something very real. Players invest hundreds of hours of their time in their game per year and many of these games request the player to pay a subscription fee of about 15 dollars / Euros a month. This means that a player who has played a game for a year or so really has invested a considerable amount of real-life effort in the game. It is therefore plausible to assume that many (but definitively not all) choices made by the player will reflect the choices he or she would make in the real world. Some actions will not reflect real-world action; the player could be an active role-player. In this case some of his or her actions do not reflect the actions taken in the real world. However at a more fundamental level this difference remains to be seen: even a role-player will not buy items at a higher price if these are available at a lower price.

An interesting third indication of the seriousness of MMOGs is the concept and consequences of death [7]. Depending on how death is implemented in the MMOG, the consequences are very different. If death means loss of equipped items, player-versus-player competition in battles is a risky business (EVE-online). If death means being relocated to a "resurrection" spot with penalties to your characters abilities, it becomes even riskier. Regardless the actual form of death, death in MMOGs is a strong mechanism to shape the game as well as how gamers make choices[7]. The fact that in-game death of a character is important to players is a strong indication that players really invest in and care about the game-world and their actions therein.

Fourth, there are strong indications that the way in which players manage their virtual teams is both a useful learning experience for as well as maps to how players manage real teams. Companies such as IBM are currently researching the links between MMOG team management and, for example, the concept of leadership [6]. Additionally, research using MMOGs as tools in the areas of inter-human trust and conflict is currently being initiated [3].

Given these strong indications that players do care about their game world and character, and that their behavior might very well be compatible with their behavior in the real world, an MMOG becomes an interesting research tool for the large-scale study of, for example, social phenomena and policy impact. Typically, hundreds of thousands of strongly involved persons play an MMOG, and the game mechanics can be set up such that specific social experiments can be investigated. Further, the investments might seem large, but this is not true. A reasonably good MMOG needs a couple of million dollars / Euros before launch, which is actually quite small an investment compared to the cost of policy making in general. If these costs are shared between governments and game developers, this might even be an interesting business model for game development (and a huge marketing perk: we develop social and environmentally sound games, we help the world, see e.g. the planned inclusion of environmental mechanisms in the new Sims game by Electronic Art).

In the rest of this paper, we explain why the current green revolution is an excellent and timely candidate for experimentation in an MMOG environment. We present an example game concept of how such an MMOG could be developed.

## 3   Energy and Pollution laws: an MMOG experiment

Although there is a strong awareness of the need to change how we are currently interacting with the world (see the recent impact of the Al Gore environmental campaign as well as books by McDonough & Braungart [8]), most governments are pretty uncertain as to how to implement "green laws" while keeping the economy up and running especially in areas such as China and India (see, e.g., [12]). Pollution and energy management are global issues, and the consequences of local laws are difficult to oversee. Consider the following two side effects of new $CO_2$ emissions laws. As soon as green energy became mass marketable (and, not unimportantly, funded such as by the Dutch government), energy companies started to invest in so called green energy production. Of course some of this investment is rock-solid green, however a lot of it is grayish green. For example, palm oil plantations function as green (as in $CO_2$ neutral) energy source, but cause damage to rainforests, and countries that produce palm oil have much less strict regulations on pollution. As a result the refineries themselves can be heavy polluters. Another side effect is that some countries now want their nuclear energy production (no $CO_2$ emission) to be considered as green. Regardless of ones opinion on these issues, they are somewhat unanticipated side effects of laws on $CO_2$ emissions.

In general, what seems to be the problem is that there is no good way to experiment with new laws [2, 12]. Especially global effects of laws are difficult to anticipate. MMOGs can play an important role. An imaginary but plausible system of laws can be implemented in an MMOG. These laws are an abstraction or simplification of existing laws and together construct the game rules in the MMOG. When the effect of a new law, let us assume a new way of taxing pollution, is to be evaluated, the MMOG can be used to as tool. The law is introduced as a new game rule, and the players will react in some way to the introduction of the new law. As MMOG's are very large-scale and players are typically scattered across the globe, a large variety of persons will interact with the law system and the new law. In terms of social experimentation this means that MMOGs have a very broad population sample. Further, they are long-term games, meaning that changes to stable game mechanisms (read: new laws) can be measured and evaluated. Players are very creative in trying to exploit and break the game's mechanism (read: players test the law system). Players are very active thinkers and participate in making the game work better (read: players are active citizens and think about the laws). To summarize, an MMOG can be used to experiment with policy making, even if the in-game laws differ from the real-world laws. An MMOG can be used to evaluate potential misuse of laws and stability issues, and thereby MMOGs can be a useful tool to anticipate the effects of new laws.

Of course players will gradually adapt to the game, which means that the game might trigger players to think about real-life energy and pollution problems. In and of itself this would be a very positive byproduct. However, this byproduct will pollute the population sample from a research perspective. This means that (a) the way the MMOG is used as a tool or (b) the type of laws that can be evaluated might change over the lifetime of the game. From a serious-gaming perspective this introduces an important research question: if large games are used as research tool, how does the population sample (gamers) influence the research and how to cope with this. We do not address this question in this paper.

In the last section of this paper we present the concept for an MMOG that focuses on experimenting with energy and pollution laws on a global scale. The imaginary (but plausible) laws are implemented in the game mechanism not in the game goals. The goal is to become successful, rich and important (as in many MMORPGs). The mechanism ensures this can only be done by caring for the environment or exploiting the rules.

## 4   *Energetics*: a Realistic Environmental MMOG

*Energetics* is a multiplayer online game in the same spirit as *EVE Online*, *World of Warcraft* or *Everquest*. It presents the player with a simulated fantasy world inhabited by other players. The world has natural resources of different types. The game setting is a post-industrial society with a strong top-down (governmental) force to reduce energy consumption for two reasons. First, fossil fuels will run out soon, and the world's countries have had great difficulty regulating fossil fuel use, as well as developing economically viable alternative energy sources. Second, energy production and therefore energy use is directly related to pollution, a major problem in many of the world's regions. Therefore, most governments in the game world have declared that energy is the major asset of our world, and decided to ration all citizens to a limited amount of energy. The chosen mechanism is progressive energy tax combined with a player-based fixed cap: the more you use above your cap, the more you pay proportionally (please note that taxation is mentioned as one of the most promising mechanism testable in MMOGs [2]). The main challenge in the game is to accumulate wealth and power. In order to do so, players can buy, produce and sell raw materials, knowledge, pollution quota and energy quota. In addition, players can earn money by producing energy themselves. All energy they produce can be sold. Players can organize themselves freely and different forms of corporations are possible. Governments sometimes oppose new regulations and rules, and it is therefore the player's goal to anticipate and adapt to these without losing profit.

### 4.1   Game concepts

Everything in *Energetics* resolves around profit, production, trade, energy, pollution, knowledge, innovation, and quota. We will explain these game concepts in detail.

People want to make profit, or at least level income versus outcome. This is true for most people in almost all regions of the world in all times. In *Energetics*, profit can be a result of production or trade. Production (energy, items and knowledge) produces pollution and needs energy. Every player or corporation has tradable pollution and energy quota. A player (or player corporation) can produce in addition to knowledge and energy, a variety of items that can potentially be useful to other players.

### 4.2 Energy production and trade

Energy can be produced by using a wide range of energy generator types, such as refineries, plutonium refiners, solar cell blocks, water power generators, wind energy, tide power generators, nuclear energy, fossil fuel energy plants, and biomass energy plants. Energy plants can produce different types of energy, such as electricity, heat, fossil fuel and nuclear fuel. Combinations of these plants should be possible but this depends on the player or corporation knowledge as well as skills. Energy plants as well as other structures need to be built, which (for the larger ones) costs a lot of items and natural resources including energy. All resources and items have the exact amount of energy they have cost to produce associated with them. All energy cost have been incorporated, such as research, production and transport. As there is strong taxation on energy use, it is beneficial for overall profit to use the materials that have the lowest associated energy cost. The building process itself also costs energy, mostly in the form of fuel and electricity for the construction yard. As players and corporations are taxed based on their energy use, this means that building, for example, a power plant is a substantial investment. In addition to that, production of energy may cost energy as well (e.g., when fossil fuel or biomass is used to generate electricity) and always produces pollution. The amount of pollution produced by the plant is added to the total for the owner (player or corporation). If the owner exceeds his quota it has to choose to shut down activities (e.g. a plant or item factory owned). Energy can be sold to other players directly, as well to the national or global grid. The amount of energy in the grid is tradable as a resource. Energy units have an associated percentage of tax that is dependent on the amount already bought (progressive taxing). So, the more energy a player buys, the large the proportion of tax. World governments have decided to produce a certain amount of energy themselves, so there is always some energy in the global grid (for balancing and bootstrapping).

### 4.3 Item production and trade

Items can be produced in the same way as energy. A production facility must be build. This costs energy, items and resources and produces pollution. Pollution is added to the owner's pollution use. The total amount of energy used to develop the factory is then used to determine the amount of energy associated with each item. This depends on the amount of items the owner estimates to build. For example, if a total of 1000MWh (Mega Watt Hour) has been used to build the factory, and the owner estimates to build 1000 items, then the associated energy cost for one item equals 1MWh plus the production cost of that item, let's say 0.5MWh. The total would then be 1.5MWh. Again, items can be traded between players (corporations) or through the national and global market. As any item has an associated real energy cost, players not only buy an item they also buy energy usage. This item-specific energy use is added to the buyers energy use when the item is bought. This means that it is progressively expensive to buy items, as every item adds to the total of the buyers energy use.

### 4.3 Knowledge, innovation and trade

Players and corporations can choose to invest in knowledge production instead of energy or item production. Knowledge enables development of better, cleaner and faster production facilities and energy plants. It also enables the combination of different energy plants (e.g., combined fossil fuel and biomass) and the combination of production mechanisms to produce more advanced items. Investing in knowledge is cheap in terms of energy usage, but expensive in terms of time and money. Knowledge requires the building of a research facility. Again, building the facility costs energy, resources, and items and produces pollution. Pollution is added to the owner's pollution use, while energy use to build the facility is distributed among the number of items (in this case knowledge licenses). In contrast to item and energy production, research needs actual player input and time. A research facility thus enables players to privately discuss new mechanisms, test out new mechanisms, build prototypes etc.. The game is set up in such a way that players can change portions of the game mechanics to find out better ways of tweaking energy and items production. Once players come up with a new mechanism, they have to file it to the Energetic patent bureau (a player driven neutral organization), that will investigate if the mechanism is doable within the game, has enough impact to be patentable and does not destabilize energy, pollution and market. The real-world play time it took to come up with the innovation of the players involved is calculated into an Energetic energy measure and pollution output. This is the amount of innovation specific energy cost, which will be added to the innovation-specific portion of the energy cost of the facility. This total is distributed over the number of licenses planned by the player or corporation that is the owner of the innovation. This links the real and virtual world in a new way and favors critical players. The innovation can be sold (again incorporating energy cost) to other players and corporations either directly, nationally or globally.

### 4.4 Energy and pollution quota

The game resolves around energy use and pollution. Every player has a fixed amount of pollution he/she can produce for free per week. This is equal for all. If a player or corporation exceeds its maximum allowed consumption, a forced shutdown of one or more activities is triggered. The player has to shutdown those activities that would produce more pollution in the current week than allowed. This can thus severely hinder profit making. However, pollution quota can be increased by buying pollution rights (quota) from other players who have not used their total amount last week. Pollution rights are therefore a tradable resource (there is a cap and trade mechanism for pollution). The same counts for energy rights, but trading these works a little different. As players and corporations can buy as much energy as needed, assuming they are willing to pay the associated progressive tax, there is no hard cap on energy consumption. However, to enable less wealthy individuals to undertake normal living activities, every government has defined a maximum amount of energy a player can use without taxation. If, for some reason, a player or corporation did not use this tax-free amount in full, it can choose to sell it to others, again directly or through the different markets. This amount can thus be bought by players and corporations who are in need of energy but heavily exceed their allowed tax-free amount (a mixture of cap and trade and taxation is used, see [12]).

### 4.5  Items, knowledge and energy production related quota

The impact of buying items and licenses and using energy is straightforward. Energy associated with the item, license or energy unit is added to the player's or corporation total for this week. (remember: an energy unit can have a higher energy value than it actually represents in pure energy due to the energy needed to built the plant and potential use of fossil or nuclear power). There is however one exception. The amount of associated energy depends on the number of units produced (again due to the energy investment of the facility), and therefore is an estimate. If for some reason the plant stops producing the units (licenses, items or energy), the estimate was too low: an energy dept exists. This dept represents the difference between the energy investment needed to build the facility and the total transferred energy quota to other players or corporations through sold units. For example if an owner used 1000MWh to built a plant, and started selling units based on an estimate of a total of 1000 units over the facility's lifetime, every unit sold had a 1MWh energy quota associated with it. If only 500 units get sold after which the facility is put out of service, the owner has an energy dept of 500MWh. The facilities owner is the first responsible to pay this dept. This means that the owner has to compensate this dept. This can be done in 3 ways. First, the owner buys the same amount of energy off the market (including progressive taxation) to compensate for it (meaning in the example that the owner has to buy 500MWh of the market). This compensation cannot be used by the owner. It is instead donated to the governments that use it for their default energy production. Second, the owner can try to sell the dept to players and corporations directly or through the markets. Players that still have tax-free energy left could be interested. In this case, the dept is added to the interested party's energy quota for this week (probably accompanied with a large sum of money). Third, a player or corporation can choose to compensate the dept by temporarily lowering its tax-free amount of allowed energy per week until the amount saved equals the dept.

### 4.6  Transport

Items, resources and energy need to be transported. Different transport facilities exist (think of power cables, trucks, ships, etc.). A transport facility is actually a mobile production facility that takes in items at one place and spits them out at another. Therefore, all production facility rules apply to transport too. For example, if you decide to build a cargo ship, this costs energy and pollution. This facility takes in items or resources, uses energy and produces pollution to reproduce the same items and resources at a different site. Again, the items and resources get transferred a proportion of the energy investment of the transport facility energy build cost. The amount of energy used and pollution produced is dependent on the length of the journey, the efficiency of the facility and the cargo weight.

### 4.7 Virtual-world specifics

There are location-dependent settings in the world of *Energetics*. First, not all countries have the same taxation and quota schemes, although they all share the same principles: progressive energy taxation and cap-based pollution quota. Second, facilities (items, energy, or knowledge) use or produce a different amount of units in different places. As transport of resource materials costs a lot (money, energy, pollution), these two location-based differences ensure an interesting tradeoff between building a facility where resources are abundant and shipping those versus building the facility at a less favorable location. This also means that trading can be a profitable way of making money. On the other hand, small-scale local production initiatives that produce for local markets tend to have low energy cost associated (no transport, small-scale thus low energy use thus low-taxation, etc.) and thus cheap products. Third, grouping together can create very large initiatives that have huge buffers of pollution and energy rights, as the quota of individual players can be transferred (0-100%) from the player's quota to the corporation quota. Fourth, governments will from time to time change their taxation, licensing and pollution laws (a new policy experiment). The way players adapt to these changes is input for the policy makers.

## 5 Conclusion

We have argued that MMOGs (MMORPGs, the role-player variant) are interesting research tools for policy makers. Such games involve many players, from a large number of different countries, with a strong involvement in the game. Their actions can be considered comparable with the their actions in the real-world, and the MMOGs can be developed such that the game mechanisms embed laws and rules that need to be tested without compromising the fun aspect of the game. We have presented an example of a game concept, called *Energetics*. The game obviously has flaws and is incomplete, and should be seen as an inspiration for an MMOG to test energy and pollution laws. As a potential byproduct, players might become more aware of energy and pollution related problems in real life.

**Acknowledgments.** The author would like to thank Mettina Veenstra for useful criticism and comments.